\documentclass[twocolumn,aps,pre,showpacs,preprint,amsmath,amssymb,showkeys,10pt]{revtex4-1} 

\usepackage[french,english]{babel}
\usepackage[T1]{fontenc}
\usepackage[utf8]{inputenc}
\usepackage{graphicx}
\usepackage{dsfont}
\usepackage{mathrsfs}
\usepackage{mathtools}
\usepackage[captionskip=-160pt]{subfig}
\usepackage{array}
\usepackage{hyperref}
\usepackage[scientific-notation=true]{siunitx}
\usepackage{tabularx}
\usepackage{booktabs}

\newcommand{\abs}[1]{\left\vert #1 \right\vert}

\newcommand{\vect}[1]{\boldsymbol{#1}}
\newcommand{\mtx}[1]{\boldsymbol{#1}}
\newcommand{\Tr}{\textrm{Tr}}
\newcommand{\grad}{\vect{\nabla}}
\newcommand{\dv}{{\rm div}}

\newcommand{\pj}{\mtx{P}}
\newcommand{\id}{\mtx{\mathrm{Id}}}
\newcommand{\rh}{\frac{r}{h}}
\newcommand{\rd}{{\rm d}}

\begin{document}

\title{Simulations of detonation waves with smoothed dissipative particle dynamics}

\author{G\'er\^ome Faure}
\affiliation{ CEA, DAM, DIF, F-91297 Arpajon, France }
\author{Jean-Bernard Maillet}
\affiliation{ CEA, DAM, DIF, F-91297 Arpajon, France }

\date{\today}

\begin{abstract}
  Smoothed Dissipative Particle Dynamics (SDPD) is a mesoscopic method which allows to select the level of resolution at which a fluid is simulated.
  The aim of this work is to extend SDPD to chemically reactive systems.
  To this end, an additional progress variable is attached to each mesoparticle and evolves according to chemical kinetics.
  This reactive SDPD model is illustrated with numerical studies of the shock-to-detonation transition in nitromethane as well as the stationary behavior of the reactive wave.
\end{abstract}

\maketitle

\section{Introduction}
\label{sec:introduction}

With the development of new architectures and massively parallel codes, Molecular Dynamics (MD) simulations have been applied to systems of increasing sizes, up to billions of atoms, and times, up to a few nanoseconds~\cite{glosli_2007,kadau_2010}.
Nevertheless MD simulations still cannot reach the time and length scales at which some complex phenomena, such as the build up of reactive waves in molecular systems, occur.
A variety of mesoscopic methods have been designed to stretch these scales by several orders of magnitude, at the expense of a decreasing predictive power compared to MD.
They generally consider fewer degrees of freedom and allow for larger timesteps since they do not need to track the intramolecular vibrations and use softer potentials.

Smoothed Dissipative Particle Dynamics (SDPD)~\cite{espanol_2003} has been introduced as a top-down coarse-graining approach that combines Smoothed Particle Hydrodynamics (SPH)~\cite{lucy_1977,monaghan_1977}, a particle Lagrangian discretization of the Navier-Stokes equations, and thermal fluctuations of mesoscopic models such as Dissipative Particle Dynamics with Energy conservation (DPDE)~\cite{avalos_1997,espanol_1997}.
This makes the model thermodynamially suitable to study hydrodynamics at nanoscale.
Besides it has been shown to give results consistent with MD for a wide range of resolutions, at equilibrium and for shock waves~\cite{faure_2016}, as well as dynamical properties such as the diffusion coefficient of a colloid in a SDPD bath~\cite{vazquez_2009,litvinov_2009}.
In particular SDPD has been used to study colloids~\cite{vazquez_2009,bian_2012} or polymer suspensions~\cite{litvinov_2008}. 

Detonation waves have been simulated successfully with MD for model systems~\cite{brenner_1993,rice_1996,holian_2002,heim_2007,herring_2010} in order to study the shock-to-detonation transition and the structure of the reactive waves.
The formation of detonation wave is a complex phenomenon which requires to handle chemical reactions along with the hydrodynamic behavior of the material, two processes that occur at very different time and length scales.
Due to the cost of atomistic reactive potentials, MD is limited to a rather short spatial and temporal domain while hydrodynamic method cannot provide an accurate description of chemical reactions.
It was recently proposed to use a moving window technique to study detonation waves for longer times with a limited number of atoms~\cite{zhakhovsky_2014}.

The modeling of reactive systems at a mesoscopic size is of great interest to deal with more realistic time and length scales.
DPDE~\cite{avalos_1997,espanol_1997} is such a model where one can coarse-grain a molecule into a single particle, reducing the number of degrees of freedom explicitly described.
Thanks to its energy conservation, it is able to deal with shock waves~\cite{stoltz_2006} and it has also been extended to reactive materials~\cite{maillet_2007,brennan_2014}.
Reactive DPDE has proved to give insight on the shock-to-detonation transition for nitromethane~\cite{maillet_2011}.
Reactive mechanisms have also been included in other coarse-grained dynamics~\cite{antillon_2014} and in SPH for the simulation of detonation wave~\cite{yang_2013}.

We adapt the mechanisms proposed in~\cite{maillet_2007} for DPDE to the SDPD setting.
We aim to illustrate the ability of SDPD to gain access to scales MD cannot reach while retaining some of its features like consistent thermodynamic fluctuations.
This article is organized as follows.
We first present in Section~\ref{sec:equations} the equations of SDPD as reformulated in~\cite{faure_2016}.
In Section~\ref{sec:chemistry}, we introduce our reactive mechanism for SDPD which is inspired by the work of~\cite{maillet_2007} for DPDE.
We illustrate the reactive SDPD model with the simulation of detonation wave for nitromethane in Section~\ref{sec:results}.

\section{Smoothed Dissipative Particle Dynamics}
\label{sec:equations}

At the hydrodynamic scale, the dynamics of the fluid is governed by the Navier-Stokes equations~\eqref{eq:navier-stokes}, which read in their Lagrangian form when the heat conduction is neglected (for time $t\geq0$ and position $\vect{x}$ in a domain $\Omega\subset \mathbb{R}^3$):
\begin{equation}
  \label{eq:navier-stokes}
  \begin{aligned}
    {\rm D}_t\rho + \rho\,\dv_{\vect{x}}\vect{v} &= 0,\\
    \rho {\rm D}_t\vect{v} &= \dv_{\vect{x}}\left(\mtx{\sigma}\right),\\
    \rho{\rm D}_t\left(u + \frac12\vect{v}^2\right) &= \dv_{\vect{x}}\left(\mtx{\sigma}\vect{v}\right).
  \end{aligned}
\end{equation}
The material derivative used in the Lagrangian description is defined as
\[
  D_t f(t,\vect{x}) = \partial_t f(t,\vect{x}) + \vect{v}(t,\vect{x})\grad_{\vect{x}}f(t,\vect{x}).
\]
The unknowns are $\rho(t,\vect{x}) \in \mathbb{R}$ the density of the fluid, $\vect{v}(t,\vect{x}) \in \mathbb{R}^3$ its velocity, $u(t,\vect{x}) \in \mathbb{R}$ its internal energy and $\mtx{\sigma}(t,\vect{x}) \in \mathbb{R}^{3\times 3}$ the stress tensor:
\begin{equation}
\label{eq:stress-tensor}
  \mtx{\sigma} = P\id + \eta(\grad \vect{v} + (\grad\vect{v})^T) + \left(\zeta-\frac23\eta\right)\dv(\vect{v})\id,
\end{equation}
where $P$ is the pressure of the fluid, $\eta$ the shear viscosity and $\zeta$ the bulk viscosity.

In the following, we first present the principles of the particle discretization of the Navier-Stokes equations with SPH in Section~\ref{sec:sph}.
We then introduce in Section~\ref{sec:sdpd} the SDPD equations reformulated in terms of internal energies~\cite{faure_2016}.

\subsection{Particle discretization}
\label{sec:sph}

Smoothed Particle Hydrodynamics~\cite{lucy_1977,monaghan_1977} is a Lagrangian discretization of the Navier-Stokes equations~(\ref{eq:navier-stokes}) on a finite number $N$ of fluid particles playing the role of interpolation nodes.
These fluid particles are associated with a portion of fluid of mass $m$.
They are located at positions $\vect{q}_i \in \Omega$ and have a momentum $\vect{p}_i \in\mathbb{R}^{3}$.
The internal degrees of freedom are represented by an internal energy $\varepsilon_i \in \mathbb{R}$.

\subsubsection{Approximation of field variables and their gradients}
\label{sec:approx-sph}

In the SPH discretization, the field variables are approximated as the average of their values at the particle positions weighted by a smoothing kernel function $W$ with finite support.
We introduce the smoothing length $h$ defined such that $W(\vect{r})=0$ if $\abs{\vect{r}} \geq h $.
In the sequel, we use the notation $r = \abs{\vect{r}}$.
In this work, we rely on a cubic spline~\cite{liu_2003}, whose expression reads
\begin{equation}
  \label{eq:sdpd-cubic-w}
  W(\vect{r}) = \left\{
    \begin{array}{cl}
      \displaystyle \frac{8}{\pi h^3} \left(1-6\frac{r^2}{h^2}+6\frac{r^3}{h^3}\right) & \displaystyle \text{ if } r \leq \frac{h}{2},\\[1em]
      \displaystyle \frac{16}{\pi h^3} \left(1-\frac{r}{h}\right)^3 & \displaystyle \text{ if } \frac{h}{2} \leq r \leq h,\\[1em]
      0 & \displaystyle  \text{ if } r \geq h.
    \end{array}
  \right.
\end{equation}
The field variables are then approximated as
\begin{equation}
  \label{eq:sph-approx}
  f(\vect{x}) \approx \sum_{i=1}^N f_i W(\vect{x}-\vect{q}_i),
\end{equation}
where $f_i$ denotes the value of the field $f$ on the particle~$i$.

The approximation of the gradient $\grad_{\vect{x}} f$ is obtained by deriving equation~\eqref{eq:sph-approx}, which yields
\[
  \grad_{\vect{x}} f(\vect{x}) \approx  \sum_{i=1}^N f_i \grad_{\vect{x}}W(\abs{\vect{x}-\vect{q}_i}).
\]
In order to have more explicit expressions, we introduce the function $F$ such that $\vect{\nabla}_{\vect{r}} W(\vect{r}) = -F(\abs{\vect{r}})\vect{r}$, which in the case of the cubic spline~(\ref{eq:sdpd-cubic-w}) is given by
\[
  F(r) = \left\{
    \begin{array}{cl}
      \displaystyle \frac{48}{\pi h^5} \left(2-3\rh\right) & \displaystyle \text{ if } r \leq \frac{h}{2},\\[1em]
      \displaystyle \frac{48}{\pi h^5} \frac1{r} \left(1-\rh\right)^2& \displaystyle \text{ if } \frac{h}{2} \leq r \leq h,\\[1em]
      0 & \displaystyle \text{ if } r \geq h.
    \end{array}
    \right.
\]
The expression of the approximated gradient finally becomes
\[
  \grad_{\vect{x}} f(\vect{x}) \approx  -\sum_{i=1}^N f_i F(\abs{\vect{x}-\vect{q}_i})(\vect{x}-\vect{q}_i).
\]

In order to simplify the notation, we define the following quantities for two particles $i$ and $j$:
\[
  \vect{r}_{ij} = \vect{q}_i - \vect{q}_j,\quad
  r_{ij} = \abs{\vect{r}_{ij}},\quad
  \vect{e}_{ij} = \frac{\vect{r}_{ij}}{r_{ij}},\quad
  F_{ij} = F(r_{ij}).
\]
We can associate a density $\rho_i$ and volume $\mathcal{V}_i$ to each particle as
\begin{equation}
  \label{eq:sdpd-rho-v}
  \rho_i(\vect{q}) = \sum_{j=1}^N mW(\vect{r}_{ij}),\quad
  \mathcal{V}_i(\vect{q}) = \frac{m}{\rho_i(\vect{q})}.
\end{equation}
The corresponding approximations of the density gradient evaluated at the particle points read
\begin{equation}
  \label{eq:gradient-rho}
  \grad_{\vect{q}_j} \rho_i = \left\{
    \begin{array}{cl}
      m F_{ij}\vect{r}_{ij} & \text{ if } j\neq i,\\[.5em]
      -m \sum\limits_{j=1}^N F_{ij}\vect{r}_{ij} & \text{ if } j=i.
    \end{array}
  \right.
\end{equation}

The smoothing length needs to be adapted to the size of the SDPD particles, defined as $\displaystyle K=\frac{m}{m_0}$ with $m_0$ the mass of a single microscopic particle (typically a molecule).
This is essential for the approximations~(\ref{eq:sph-approx}) to remain meaningful.
In order to keep the average number of neighbors roughly constant in the smoothing sum, we associate a smoothing length $h_K$ for each particle size $K$ with
\[
  h_K = \mathfrak{h}\left(\frac{m_K}{\rho}\right)^{\frac13}.
\]
In this work, we have taken $\mathfrak{h}=2.5$, which correspond to a typical number of 60-70 neighbors, a commonly accepted number~\cite{liu_2003}.

\subsubsection{Thermodynamic closure}
\label{sec:thermo-closure}
As in Navier-Stokes hydrodynamics, an equation of state is required to close the set of equations provided by the SPH discretization.
This equation of state relates the entropy $S_i$ of the mesoparticle $i$ with its density $\rho_i(\vect{q})$ (as defined by~\eqref{eq:sdpd-rho-v}) and its internal energy $\varepsilon_i$ through an entropy function
\begin{equation}
  \label{eq:sdpd-eos}
  S_i(\varepsilon_i,\vect{q})=\mathcal{S}(\varepsilon_i,\rho_i(\vect{q})).
\end{equation}
It is then possible to assign to each particle a temperature
\[
  T_i(\varepsilon_i,\vect{q}) = \left[\frac1{\partial_{\varepsilon}\mathcal{S}}\right](\varepsilon_i,\rho(\vect{q})),
\]
pressure
\[
  P_i(\varepsilon_i,\vect{q}) = -\frac{\rho(\vect{q})^2}{m}\left[\frac{\partial_{\rho}\mathcal{S}}{\partial_{\varepsilon}\mathcal{S}}\right](\varepsilon_i,\rho(\vect{q})),
\]
and heat capacity at constant volume
\[
  C_{v,i}(\varepsilon_i,\vect{q}) = -\left[\frac{(\partial_{\varepsilon} \mathcal{S})^2}{\partial_{\varepsilon}^2\mathcal{S}}\right](\varepsilon_i,\rho(\vect{q})).
\]
To simplify the notation, we omit in Sections~\ref{sec:eom-sdpd} the dependence of $T_i$, $P_i$ and $C_{v,i}$ on the variables $\varepsilon_i$ and $\vect{q}$.

\subsection{Equations of motion for SDPD}
\label{sec:sdpd}
Smoothed Dissipative Particle Dynamics~\cite{espanol_2003} is a top-down mesoscopic method relying on the SPH discretization of the Navier-Stokes equations with the addition of thermal fluctuations which are modeled by a stochastic force.
In its reformulated form~\cite{faure_2016}, SDPD is a set of stochastic differential equations for the following variables: the positions $\vect{q}_i\in\Omega\subset\mathbb{R}^{3}$, the momenta $\vect{p}_i\in\mathbb{R}^{3}$ and the energies $\varepsilon_i\in \mathbb{R}$ for $i=1\dots N$.

The dynamics can be split into two elementary dynamics, the first one being a conservative dynamics derived from the pressure gradient in the stress tensor~\eqref{eq:stress-tensor} and the second a set of pairwise fluctuation and dissipation dynamics stemming from the viscous terms in~\eqref{eq:stress-tensor} coupled with random fluctuations.

\subsubsection{Conservative forces}
\label{sec:conservative-sdpd}
The elementary force between particles $i$ and $j$ arising from the discretization of the pressure gradient in the Navier-Stokes momentum equation reads
\begin{equation}
  \label{eq:cons-forces}
  \vect{\mathcal{F}}_{{\rm cons},ij} = m^2\left(\frac{P_i}{\rho_i^2}+\frac{P_j}{\rho_j^2}\right)F_{ij}\vect{r}_{ij}.
\end{equation}
This part of the dynamics preserves the entropies $S_i$ along with the total energy
\[
  E(\vect{q},\vect{p},\varepsilon) = \sum_{i=1}^N \varepsilon_i + \frac{\vect{p}_i^2}{2m}.
\]
As a consequence, the variation of the internal energy only emerges from the variation of the particle volume as
\[
  \begin{aligned}
    \rd\varepsilon_i &= - P_i\rd\mathcal{V}_i,\\
    &= -\sum_{j\neq i}\frac{m^2P_i}{\rho_i(\vect{q})^2}F_{ij}\vect{r}_{ij}\cdot\vect{v}_{ij}\,\rd t.
  \end{aligned}
\]
This allows us to write the conservative part of the dynamics as
\begin{equation}
  \label{eq:sdpd-cons}
  \left\{\begin{aligned}
      \rd\vect{q}_i &= \frac{\vect{p}_i}{m}\,\rd t,\\
      \rd\vect{p}_i &= \sum_{j\neq i} \vect{\mathcal{F}}_{{\rm cons},ij}\,\rd t,\\
      \rd\varepsilon_i &= -\sum_{j\neq i}\frac{m^2P_i}{\rho_i(\vect{q})^2}F_{ij}\vect{r}_{ij}\cdot\vect{v}_{ij}\,\rd t.
  \end{aligned}\right.
\end{equation}

\subsubsection{Fluctuation and Dissipation}
\label{sec:fd-sdpd}
The viscous term in the Navier-Stokes equations translates into a dissipative force in the equation of motions.
This term is coupled with a fluctuation force that distinguishes SDPD from a mere particular discretiztion of Navier-Stokes like SPH.
In order to give the expression of the viscous and fluctuating part of the dynamics, we define the relative velocity for a pair of particles $i$ and $j$ as
\[
\vect{v}_{ij} = \frac{\vect{p}_i}{m}-\frac{\vect{p}_j}{m}.
\]

In the spirit of DPDE, we choose a pairwise fluctuation and dissipation term for $i<j$ of the following form
\begin{equation}
  \label{eq:sdpd-simple-fluct}
  \left\{
  \begin{aligned}
    \rd\vect{p}_i &= -\mtx{\Gamma}_{ij}\vect{v}_{ij}\,\rd t + \mtx{\Sigma}_{ij}\rd\vect{B}_{ij},\\
    \rd\vect{p}_j &= \mtx{\Gamma}_{ij}\vect{v}_{ij}\,\rd t - \mtx{\Sigma}_{ij}\rd\vect{B}_{ij},\\
    \rd\varepsilon_i &= \frac12\left[\vect{v}_{ij}^T\mtx{\Gamma}_{ij}\vect{v}_{ij} - \frac{\Tr(\mtx{\Sigma}_{ij}\mtx{\Sigma}_{ij}^T)}{m}\right]\rd t -\frac12 \vect{v}_{ij}^T\mtx{\Sigma}_{ij}\rd\vect{B}_{ij},\\
    \rd\varepsilon_j &= \frac12\left[\vect{v}_{ij}^T\mtx{\Gamma}_{ij}\vect{v}_{ij} - \frac{\Tr(\mtx{\Sigma}_{ij}\mtx{\Sigma}_{ij}^T)}{m}\right]\rd t -\frac12 \vect{v}_{ij}^T\mtx{\Sigma}_{ij}\rd\vect{B}_{ij},
  \end{aligned}
  \right.
\end{equation}
where $\vect{B}_{ij}$ is a $3$-dimensional vector of standard Brownian motions, $\mtx{\Gamma}_{ij}$ and $\mtx{\Sigma}_{ij}$ are $3\times3$ symmetric matrices.
In the dynamics~\eqref{eq:sdpd-simple-fluct}, the equations acting on the momenta preserve the total momentum in the system.
Furthermore, as in DPDE, the equations for the energy variables are determined to ensure the conservation of the total energy $E(\vect{q},\vect{p},\varepsilon)$.
Since $\displaystyle \rd \varepsilon_i = -\frac12 \rd \left(\frac{\vect{p}_i^2}{2m} + \frac{\vect{p}_j^2}{2m}\right)$, It\^o calculus yields the resulting equations in~\eqref{eq:sdpd-simple-fluct}.

We consider friction and fluctuation matrices of the form
\begin{equation}
  \label{eq:fluct-gamma}
  \begin{aligned}
    \mtx{\Gamma}_{ij} &= \gamma^{\parallel}_{ij}\pj^{\parallel}_{ij} + \gamma^{\perp}_{ij}\pj^{\perp}_{ij},\\
    \mtx{\Sigma}_{ij} &= \sigma^{\parallel}_{ij}\pj^{\parallel}_{ij} + \sigma^{\perp}_{ij}\pj^{\perp}_{ij},
  \end{aligned}
\end{equation}
with the projection matrices $\pj^{\parallel}_{ij}$ and $\pj^{\perp}_{ij}$ given by
\[
  \pj_{ij}^{\parallel} = \vect{e}_{ij}\otimes\vect{e}_{ij},\quad
  \pj_{ij}^{\perp} = \id - \pj_{ij}^{\parallel}
\]
Introducing the coefficients
\[
  \begin{aligned}
    a_{ij} &=\left(\frac{5\eta}{3}-\zeta\right)\frac{m^2F_{ij}}{\rho_i\rho_j},\\
    b_{ij}+\frac{a_{ij}}3 &= 5\left(\frac{\eta}3+\zeta\right)\frac{m^2F_{ij}}{\rho_i\rho_j}, \\
    d_{ij} &= k_{\rm B}\frac{T_iT_j}{(T_i+T_j)^2}\left(\frac1{C_{v,i}}+\frac1{C_{v,j}}\right).
  \end{aligned}
\]
defined from the fluid viscosities $\eta$ and $\zeta$ appearing in the stress tensor~(\ref{eq:stress-tensor}), a possible choice for the friction and fluctuation coefficients is
\begin{equation}
  \label{eq:sdpd-gamma-sigma}
  \begin{aligned}
    \gamma_{ij}^{\parallel} &= \left(\frac43a_{ij}+b_{ij}\right) \left( 1 - d_{ij}\right),\\
    \gamma_{ij}^{\perp} &= a_{ij}\left( 1 - d_{ij} \right),\\
    \sigma_{ij}^{\theta} &= 2\sqrt{\frac{\gamma_{\theta}}{1-d_{ij}} k_{\rm B}\frac{T_iT_j}{T_i+T_j}}.
  \end{aligned}
\end{equation}
This ensures that measures of the form
\begin{equation}
  \label{eq:sdpd-energy-minv}
  \begin{aligned}
    &\mu(\rd\vect{q}\,\rd\vect{p}\,\rd \varepsilon)\\
    &\,= g\left(E(\vect{q},\vect{p},\varepsilon),\sum\limits_{i=1}^N\vect{p}_i\right)\prod_{i=1}^N\frac{\exp\left(\frac{S_i(\varepsilon_i,\vect{q})}{k_{\rm B}}\right)}{T_i(\varepsilon_i,\vect{q})}\,\rd\vect{q}\,\rd\vect{p}\,\rd \varepsilon
  \end{aligned}
\end{equation}
are left invariant by the elementary dynamics~\eqref{eq:sdpd-simple-fluct} as shown is~\cite{faure_2016}.
While other forms of these coefficients are possible (for instance constant $\sigma$ parameters), the relations~(\ref{eq:sdpd-gamma-sigma}) allow to retrieve the same dissipation as in the original SPDP~\cite{espanol_2003}.

\subsubsection{Complete equations of motion}
\label{sec:eom-sdpd}
As a result, the complete set of equations of motion for SDPD reformulated in the position, momentum and internal energy variables read
\begin{equation}
  \label{eq:sdpd-energy}
  \left\{
  \begin{aligned}
    \rd\vect{q}_i =&\, \frac{\vect{p}_i}{m}\,\rd t,\\
    \rd\vect{p}_i =& \sum_{j\neq i} m^2\left(\frac{P_i}{\rho_i^2}+\frac{P_j}{\rho_j^2}\right)F_{ij}\vect{r}_{ij}\,\rd t - \mtx{\Gamma}_{ij}\vect{v}_{ij}\,\rd t\\
    &+ \mtx{\Sigma}_{ij}\rd\vect{B}_{ij},\\
    \rd \varepsilon_i =& \sum_{j\neq i} -\frac{m^2P_i}{\rho_i^2}F_{ij}\vect{r}_{ij}^T\vect{v}_{ij}\,\rd t\\
    &+ \frac12 \left[\vect{v}_{ij}^T\mtx{\Sigma}_{ij}\vect{v}_{ij} -\frac1{m}\Tr(\mtx{\Sigma}_{ij}\mtx{\Sigma}_{ij}^T)\right]\rd t\\
    & - \frac12 \vect{v}_{ij}^T\mtx{\Sigma}_{ij}\rd\vect{B}_{ij},
  \end{aligned}
  \right.
\end{equation}
where $\mtx{\Sigma}_{ij}$ and $\mtx{\Gamma}_{ij}$ are given by~\eqref{eq:fluct-gamma} and~(\ref{eq:sdpd-gamma-sigma}).
The dynamics~\eqref{eq:sdpd-energy} preserves the total momentum $\sum\limits_{i=1}^N\vect{p}_i$ and the total energy $E(\vect{q},\vect{p},\varepsilon)$ since all the elementary sub-dynamics ensure these conservations.

The time integration of the SDPD equations of motion can be performed thanks to a splitting strategy as described in~\cite{faure_2016}.
We resort to a Velocity-Verlet scheme for the conservative part given by Equation~(\ref{eq:sdpd-cons}) while for the fluctuation/dissipation part in Equation~(\ref{eq:sdpd-simple-fluct}) each pair is handled successively, following the ideas introduced for DPD in~\cite{shardlow_2003} or for DPDE in~\cite{stoltz_2006}
This scheme ensures a good energy conservation though linear energy drifts are observed in the long term.

\section{Chemical reactions}
\label{sec:chemistry}

We present the chemical mechanism included in SDPD and inspired by the reactive DPDE introduced in~\cite{maillet_2007,maillet_2011}.
We first present the modelization of the chemical kinetics in Section~\ref{sec:kinetics}.
The introduction of chemical reactions means that the material should be able to change its properties as it reacts.
This is achieved by means of a reactive equation of state, presented in Section~\ref{sec:reac-eos} that switches between the reactants and the products as the reaction occurs.
Finally, we handle the exothermicity of the reaction in Section~\ref{sec:exothermicity}.

\subsection{Kinetics of the chemical reaction}
\label{sec:kinetics}

In the spirit of~\cite{maillet_2007,maillet_2011} where chemical reactions were included in DPDE, we model the progress of chemical reactions by adding a progress variable $\lambda_i\in [0,1]$ to each mesoparticle.
Considering a model chemical reaction
\[
A \rightleftharpoons B,
\]
we associate $\lambda=0$ to the reactant $A$ and $\lambda=1$ to the product $B$.
The progress variable can be seen as the portion of the mesoparticle that has reacted.
This statistical point of view gains a clearer meaning as the size of the mesoparticle increases.

The evolution of the progress variable is governed by a kinetics that can be freely chosen to model the chemical reaction.
In this work, we adopt second order kinetics where mesoparticles can interact with neighbouring particles.
The progress variable is thus evolved as
\begin{equation}
  \label{eq:kinetics}
  \begin{aligned}
    \frac{\rd \lambda_i}{\rd t} =& \sum_{j\neq i} \mathcal{K}_{0\to1}\left(T_{ij}\right)(1-\lambda_i)(1-\lambda_j)W(r_{ij})\\
    &- \mathcal{K}_{1\to0}\left(T_{ij}\right)\lambda_i\lambda_jW(r_{ij}),
  \end{aligned}
\end{equation}
where $\mathcal{K}_{0\to1}$ and $\mathcal{K}_{1\to0}$ are the reaction rates, respectively, for the forward and backward reactions.
The reaction rates depend on the mean temperature $T_{ij} = \frac12\left(T_i+T_j\right)$ according to some Arrhenius law :
\begin{equation}
  \label{eq:arrhenius}
  \mathcal{K}_{\mathcal{X}}(T_{ij}) = Z_{\mathcal{X}}\exp\left(-\frac{E_{\mathcal{X}}}{k_{\rm B}T_{ij}}\right),
\end{equation}
with an activation energy $E_{\mathcal{X}}$, that represents the energy barrier a particle needs to overcome during the reaction, and a prefactor $Z_{\mathcal{X}}$ that governs the frequency of the reaction.
Extension of this model to several chemical reactions would be straightforward.

\subsection{Reactive equation of state}
\label{sec:reac-eos}

Since the equations of state for the reactants and for the product are different, we need to define a mixed equation of state when $0 < \lambda < 1$.
In the following, we denote all quantities related to the reactant by a superscript 0 and to the product by a superscript 1.
The functions yielding temperature and pressure from the equation of state~\eqref{eq:sdpd-eos} are thus denoted by $\mathcal{T}^0$ and $\mathcal{P}^0$ for the reactant.
The internal energy of a mesoparticle, due to its extensivity, can be expressed as
\begin{equation}
  \label{eq:mixing-ei}
  \varepsilon_i = \varepsilon_i^0 + \varepsilon_i^1,
\end{equation}
where $\varepsilon_i^0$ and $\varepsilon_i^1$ are the energies, respectively, of the reactant (a $1-\lambda$ portion of mesoparticle $i$) and the products (a $\lambda$ portion of the mesoparticle).
The density, being an intensive variable, is given as a weighted average of the density of the reactant $\rho_i^0$ and of the products $\rho_i^1$:
\begin{equation}
  \label{eq:mixing-rho}
  \rho_i = (1-\lambda)\rho_i^0 + \lambda\rho_i^1,
\end{equation}
Note that we only have access to the internal energy $\varepsilon_i$ and density $\rho_i$ (through Equation~(\ref{eq:sdpd-rho-v})) of the whole mesoparticle, along with its progress variable $\lambda_i$.
In order to obtain the temperature $T_i$, pressure $P_i$ and heat capacity $C_i$ for each mesoparticle, we need to determine the state of each chemical species ($\varepsilon_i^0$, $\rho_i^0$ and $\varepsilon_i^1$, $\rho_i^1$) thanks to a mixing law.
If $\lambda_i=0$ or $\lambda_i=1$, the mesoparticle is actually composed purely of either $A$ or $B$.
Hence we may use the equation of state for the pure chemical species.
In the other cases ($0<\lambda_i<1$), we consider the two components to be at thermal and mechanical equilibrium inside a mesoparticle, which means that
\[
  \begin{aligned}
    \mathcal{T}^0(\varepsilon_i^0,\rho_i^0) &= \mathcal{T}^1(\varepsilon_i^1,\rho_i^1), \\
    \mathcal{P}^0(\varepsilon_i^0,\rho_i^0) &= \mathcal{P}^1(\varepsilon_i^1,\rho_i^1).
  \end{aligned}
\]
Using relations~\eqref{eq:mixing-ei} and~\eqref{eq:mixing-rho} to express $\rho_i^1$ and $\varepsilon_i^1$ as a function of the global state $\varepsilon_i$, $\rho_i$ and of the state of the other component $\varepsilon_i^0$, $\rho_i^0$, this amounts to
\begin{equation}
  \label{eq:mixing-isotp}
  \begin{aligned}
    \mathcal{T}^0(\varepsilon_i^0,\rho_i^0) - \mathcal{T}^1\left(\varepsilon_i-\varepsilon_i^0,\frac{\rho_i-(1-\lambda)\rho_i^0}{\lambda}\right) &= 0, \\
    \mathcal{P}^0(\varepsilon_i^0,\rho_i^0) - \mathcal{P}^1\left(\varepsilon_i-\varepsilon_i^0,\frac{\rho_i-(1-\lambda)\rho_i^0}{\lambda}\right) &= 0.
  \end{aligned}
\end{equation}
The computation of the energy $\varepsilon_i^0$ and density $\rho_i^0$ generally requires to resort to a numerical inversion, like the Newton method, so that Equation~\eqref{eq:mixing-isotp} holds.
This finally yields the temperature $T_i=\mathcal{T}^0(\varepsilon_i^0,\rho_i^0)$ and pressure $P_i=\mathcal{P}^0(\varepsilon_i^0,\rho_i^0)$ that are used in Equation~(\ref{eq:kinetics}) for the chemical reactions and in the usual equations of motion of SDPD~(\ref{eq:sdpd-energy}).

\subsection{Exothermicity}
\label{sec:exothermicity}

Chemical reactions are called exothermic if they release some chemical energy (or heat) as they occur.
It is naturally important to take into account such effects and an exchange between the chemical energy and the other degrees of freedom occurs as the reaction progresses.
The exothermicity, which is the energy liberated by the reaction of a single molecule, is given as
\[
  E_{\rm exo} = E_{1\to0}- E_{0\to1}
\]
and the total energy in our reactive system now reads
\[
  E(\vect{q},\vect{p},\varepsilon,\lambda) = \sum_{i=1}^N \varepsilon_i + \frac{\vect{p}_i^2}{2m} + (1-\lambda_i)KE_{\rm exo}
\]
Note that the chemical energy scales with the particle size $K$.
Since we request that $E$ is exactly preserved as the reaction progresses, the exothermicity is progressively transferred in the internal energy, inducing an evolution of the internal energy given by
\[
  \rd \varepsilon_i = KE_{\rm exo}\,\rd\lambda_i.
\]
It would be possible to also release this energy in the kinetic energy at the cost of the conservation of the total momentum.
In practice the exchange of energy between the internal and external degrees of freedom quickly leads to an equilibration between the kinetic and internal energy.

This reactive mechanism is coupled with the equations of motion of SDPD~(\ref{eq:sdpd-energy}).
In order to integrate the reactive SDPD model, we use the SSA scheme described in~\cite{faure_2016}.
An additional step is included in the integration scheme where the progress variables are updated with an Explicit Euler scheme.
The internal energies are finally evolved by taking into account the exothermicity with
\[
  \varepsilon_i^{n+1} = \varepsilon_i^n + (\lambda_i^{n+1}-\lambda_i^n)KE_{\rm exo}.
\]
This ensures that the total energy is preserved when integrating the reactive part of the dynamics.

\section{Application to nitromethane}
\label{sec:results}

We assess the validity of the reactive SDPD model by simulating the propagation of a detonation wave in nitromethane.
We model the decomposition of nitromethane by a single irreversible exothermic reaction:
\[
  {\rm NiMe} \to {\rm products}.
\]
Compared to the more generic framework of Section~\ref{sec:kinetics} for reversible reactions, the irreversibility of the reaction is achieved by taking $Z_{1\to0}=0$.
The reaction rate follows the Arrhenius law specified in~\eqref{eq:arrhenius}.
The activation energy is $E_a = \num{3e-19}$~J and the exothermicity $E_{\rm exo} = \num{4.78e-19}$~J/molecule as in~\cite{maillet_2011}.
The influence of the prefactor $Z$ is investigated in Section~\ref{sec:prefactor}.

Inert nitromethane is represented by an equation of state obtained from Monte Carlo molecular simulations~\cite{desbiens_2008} with a force field optimized to reproduce the properties of nitromethane under shock~\cite{desbiens_2007}.
The analytic form of the equation of state is given by
\begin{equation}
  \mathcal{S}_{\rm NiMe}(\varepsilon,\rho) = C_{V}\log\left[\frac{\varepsilon-\mathcal{E}_{\rm ref}(\rho)}{C_{V}} + \theta(\rho)\right] + C_{V}\Gamma_0\frac{\rho_0}{\rho},
  \label{eq:eos-hz}
\end{equation}
with
\[
  \theta(\rho) = (T_0 - T_{00})\exp\left[\Gamma_0\left(1-\frac{\rho_0}{\rho}\right)\right],
\]
and
\[
  \mathcal{E}_{\rm ref}(\rho) = \frac12\frac{c_0^2x^2}{1-sx} \times \left\{
    \begin{array}{cl}
      \displaystyle 1+\frac{sx}3-s\left(\Gamma_0-s\right)\frac{x^2}6 & \displaystyle \text{ if } x \geq 0,\\
      \displaystyle 1 & \displaystyle \text{ if } x < 0,
    \end{array}
  \right.
\]
where $\displaystyle x=1-\frac{\rho_0}{\rho}$, and $T_0$ and $T_{00}$ are two constants defined as the standard temperature $T_0 = 298.13$~K and the temperature $T_{00}$ on the reference curve $\mathcal{E}_{\rm ref}$.
This constant is determined as $T_{00} = \frac{E_0}{C_V}$ where $E_0$ is the energy in standard conditions (density $\rho_0$ and pressure $P_0 = 10^5$~Pa).
The parameters of~\eqref{eq:eos-hz} are summarized in Table~\ref{tab:eos-hz}.
\begin{table}[!ht]
\centering
  \begin{tabular}{ccc}
    \toprule
    Parameter & Value & Unit\\
    \midrule
    $\Gamma_0$ & $1$ & - \\[1pt]
    $\rho_0$ & $1140$ &  kg.m$^{-3}$ \\[1pt]
    $c_0$ & $1358.47$ & m.s$^{-1}$\\[1pt]
    $C_{V}$ & $1211$ & J.K$^{-1}$.kg$^{-1}$\\[1pt]
    $s$ & $2.000184$ & - \\
    \bottomrule
  \end{tabular}
  \caption{Parameters of the equation of state~\eqref{eq:eos-hz} for nitromethane.}
  \label{tab:eos-hz}
\end{table}

The products of the reaction are modeled by a Jones-Wilkins-Lee (JWL) equation of state, introduced in~\cite{book_lee_1968} for reaction products.
It reads
\begin{equation}
  \mathcal{S}_{\rm JWL}(\varepsilon,\rho) = C_{V}\log\left[\frac{\varepsilon-\mathcal{E}_k(\rho)}{C_{V}}\right] - C_{V}\Gamma_0\log(\rho), 
  \label{eq:eos-jwl}
\end{equation}
with
\[
  \begin{aligned}
    \mathcal{E}_k(\rho) = &\frac{a}{\rho_0R_1}\exp\left[-R_1\frac{\rho_0}{\rho}\right] + \frac{b}{\rho_0R_2}\exp\left[-R_2\frac{\rho_0}{\rho}\right]\\
    &+ \frac{\mathfrak{K}}{\rho_0\Gamma_0}\left(\frac{\rho_0}{\rho}\right)^{-\Gamma_0} + C_{\rm ek},
\end{aligned}
\]
using the parameters from~\cite{book_dobratz_1985}, which are gathered in Table~\ref{tab:eos-jwl}.
\begin{table}[!ht]
  \centering
  \begin{tabular}{ccc}
    \toprule
    Parameter & Value & Unit\\
    \midrule
    $\Gamma_0$ & $0.3$ & -\\
    $\rho_0$ & $1128$ & kg.m$^{-3}$\\
    $E_0$ & $0$ & J\\
    $D_{\rm CJ}$ & $6280$ & m.s$^{-1}$\\
    $P_{\rm CJ}$ & $\num{1.25e10}$ & Pa \\
    $T_{\rm CJ}$ & $3000$ & K \\
    $C_{V}$ & $2764.23$ & J.K$^{-1}$.kg$^{-1}$\\
    $a$ & $\num{2.092e11}$ & Pa\\
    $b$ & $\num{5.689e9}$ & Pa\\
    $R_1$ & $4.4$ & -\\
    $R_2$ & $1.2$ & -\\
    \bottomrule
  \end{tabular}
  \caption{Parameters of the JWL equation of state~\eqref{eq:eos-jwl} for reacted nitromethane (products).}
  \label{tab:eos-jwl}
\end{table}
In order to define the constants $\mathfrak{K}$ and $C_{\rm ek}$, we first introduce
\[
  \left\{
    \begin{aligned}
      \rho_{\rm CJ} &= \rho_0 \frac{\rho_0D_{\rm CJ}^2}{\rho_0D_{\rm CJ}^2-P_{\rm CJ}},\\
      E_{\rm CJ} &= E_0 + \frac12P_{\rm CJ}\left(\frac1{\rho_0}-\frac1{\rho_{\rm CJ}}\right),\\
      P_{\rm k1CJ} &= a\exp\left(-R_1\frac{\rho_0}{\rho_{\rm CJ}}\right) + b\exp\left(-R_2\frac{\rho_0}{\rho_{\rm CJ}}\right).
    \end{aligned}
  \right.
\]
Then,
\[
  \mathfrak{K} = \left(P_{\rm CJ} - P_{K1{\rm CJ}} - \frac1mC_v\Gamma_0T_{\rm CJ}\rho_{\rm CJ}\right)\left(\frac{\rho_0}{\rho_{\rm CJ}}\right)^{\Gamma_0+1},
\]
and
\[
  \begin{aligned}
    C_{\rm ek} =& E_{\rm CJ} -\frac{a}{\rho_0R_1}\exp\left(-R_1\frac{\rho_0}{\rho_{\rm CJ}}\right)\\
    &-\frac{b}{\rho_0R_2}\exp\left(-R_2\frac{\rho_0}{\rho_{\rm CJ}}\right)
    -\frac{P_{\rm CJ}-P_{\rm k1CJ}}{\rho_{\rm CJ}\Gamma_0}.
  \end{aligned}
\]

In order to observe a detonation wave in our simulations, we first create a shock wave in the neat nitromethane that transforms to a detonation wave provided the shock velocity is high enough.
The time step $\Delta t$ is chosen such that the particle after the initial shock wave do not move by more than $10\%$ of the characteristic inter-particle distance during one step.
We first study in Section~\ref{sec:shock-to-deto} the transition from a shock wave to a detonation wave before turning to the analysis of the stationary behavior of the detonation wave in Section~\ref{sec:stationary}.
We conclude by studying the influence of the Arrhenius prefactor in Section~\ref{sec:prefactor}.

\subsection{Shock to detonation transition}
\label{sec:shock-to-deto}

The system we consider is formed of $N=86400$ particles initially distributed on a $12\times12\times594$ grid at the nitromethane equilibrium density $\rho = 1104$~kg.m$^{-3}$.
The initial velocities and internal energies are chosen so that the initial temperature in the system is $300$~K.
Periodic boundary conditions are used in the $x$- and $y$-directions.
In the $z$-direction, two walls, formed of ``virtual'' SDPD particles as described in~\cite{bian_2012,faure_2016}, are located at each end of the system.
These virtual particles interact with the real SDPD particles through the conservative forces~(\ref{eq:cons-forces}) and a repulsive Lennard-Jones potential that ensures the impermeability of the walls.
After the system equilibration during $\tau_{\rm therm} = 100$~ps, the lower wall is given a constant velocity $v_{\rm P} = 1764$~m.s$^{-1}$ in the $z$-direction.
We choose the viscosity parameter $\eta = \num{2e-3}$~Pa.s so that the shock profile is smooth and no spurious oscillations are observed after the shock front.
We first use a prefactor $Z=10^{15}$~s$^{-1}$ that is large enough to observe the shock-to-detonation transition in the spatio-temporal window of the simulation.

\begin{figure}
  \subfloat[]{\includegraphics{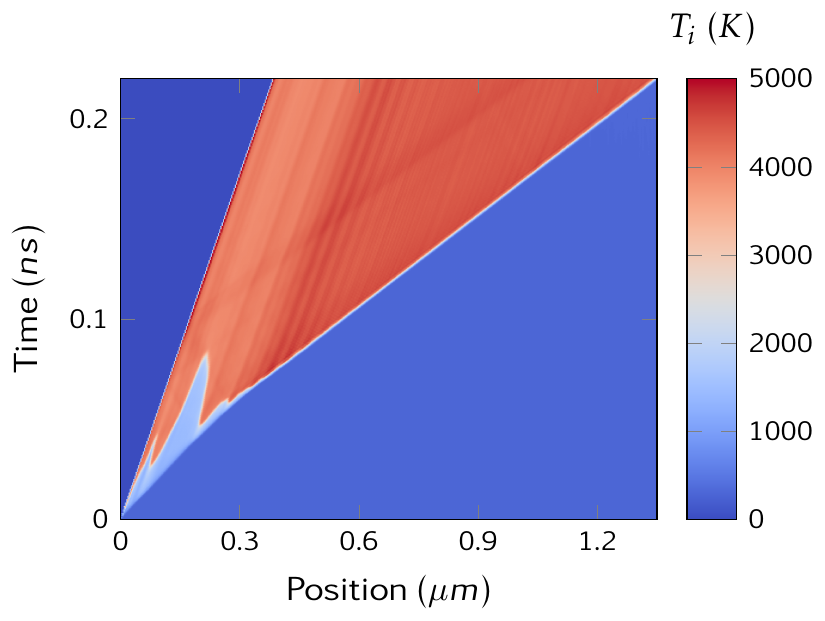}\label{fig:color-t}}\\[-1em]
  \subfloat[]{\includegraphics{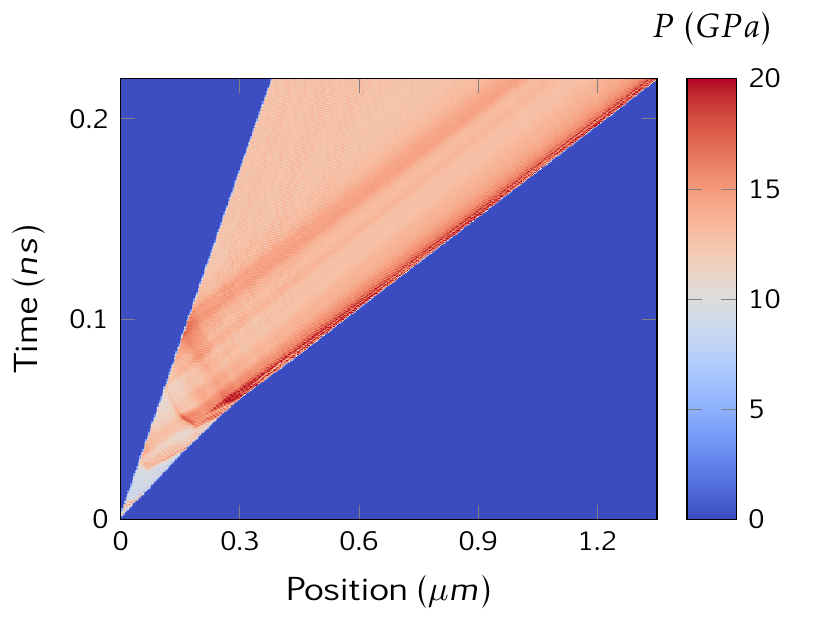}\label{fig:color-p}}\\[-1em]
  \subfloat[]{\includegraphics{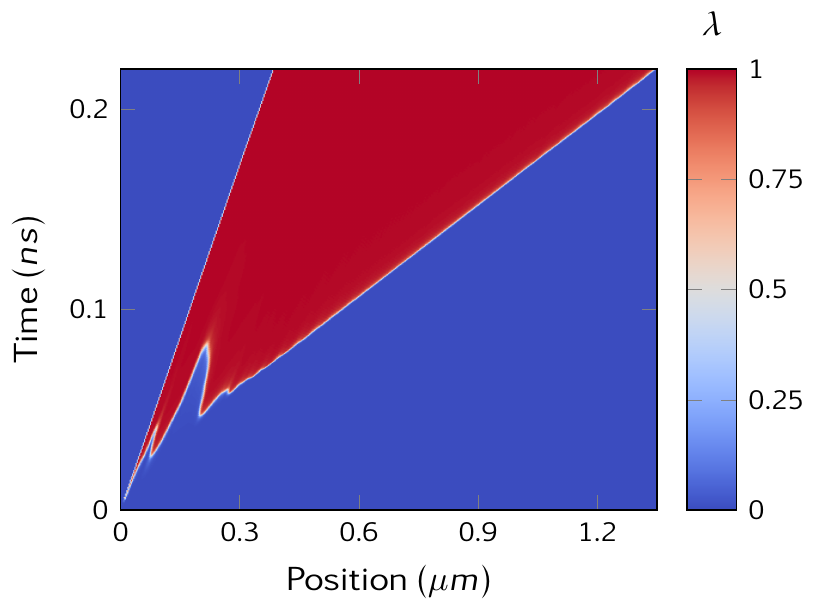}\label{fig:color-lb}}
  \caption{Space-time diagram of \protect\subref{fig:color-t}~temperature,\protect\subref{fig:color-p}~pressure and \protect\subref{fig:color-lb}~progress variables for $K=100$}
  \label{fig:color-deto}
\end{figure}

We carry several simulations for different particle sizes $K$.
Since we keep the dimensions of the system constant in reduced units, the overall size in physical units increases with the particle size.
Due to the different time and length scales, the time step is also dependent on $K$: for instance, we take $\Delta t = \num{1.3e-13}$~s for $K=10$ and $\Delta t = \num{6.0e-13}$~s for $K=1000$.
At each step, the spatial domain is split into $n_{\rm sl} = 450$ slices along the $z$-direction over which the thermodynamic variables are averaged in order to estimate instantaneous profiles.
Their evolution along time is plotted in time-space diagrams (see Figure~\ref{fig:color-deto}).
Two distinct domains can be distinguished.
First, a shock wave is formed thanks to the piston movement.
While it propagates in the material, the high temperature leads to the ignition of chemical reactions in the shocked nitromethane.
These reactions create compressive waves in the shocked material, leading to the appearance of new ignitions points, forward in the neat shocked nitromethane.
Finally as the new ignition points get closer to the shock front, they catch up with the shock front and begin to drive the shock at a larger velocity, hence forming a detonation wave which propagates at a constant velocity $v_{\rm D} = 6646$~m.s$^{-1}$.
This is in good agreement with the theoretical hydrodynamic prediction which reads $D_{\rm CJ} = 6620$~m.s$^{-1}$.
This shock-to-detonation transition mechanism is very similar to previous computations carried at a more microscopic scale with reactive DPDE~\cite{maillet_2011} where the same discontinuous process, with successive ignition points in the shocked material, was observed.
However this differs from hydrodynamic simulations where a reactive wave, first ignited close to the wall, catches up to the shock front, forming a so-called super-detonation~\cite{menikoff_2011}.
It is still not clear what is the origin of this discrepancy.
It may be an effect of the large prefactor used in~\cite{maillet_2011} and in our simulations which accelerate the chemical kinetics and change its time scale compared to the hydrodynamic time scale.

\begin{figure}[!ht]
  \centering
  \includegraphics{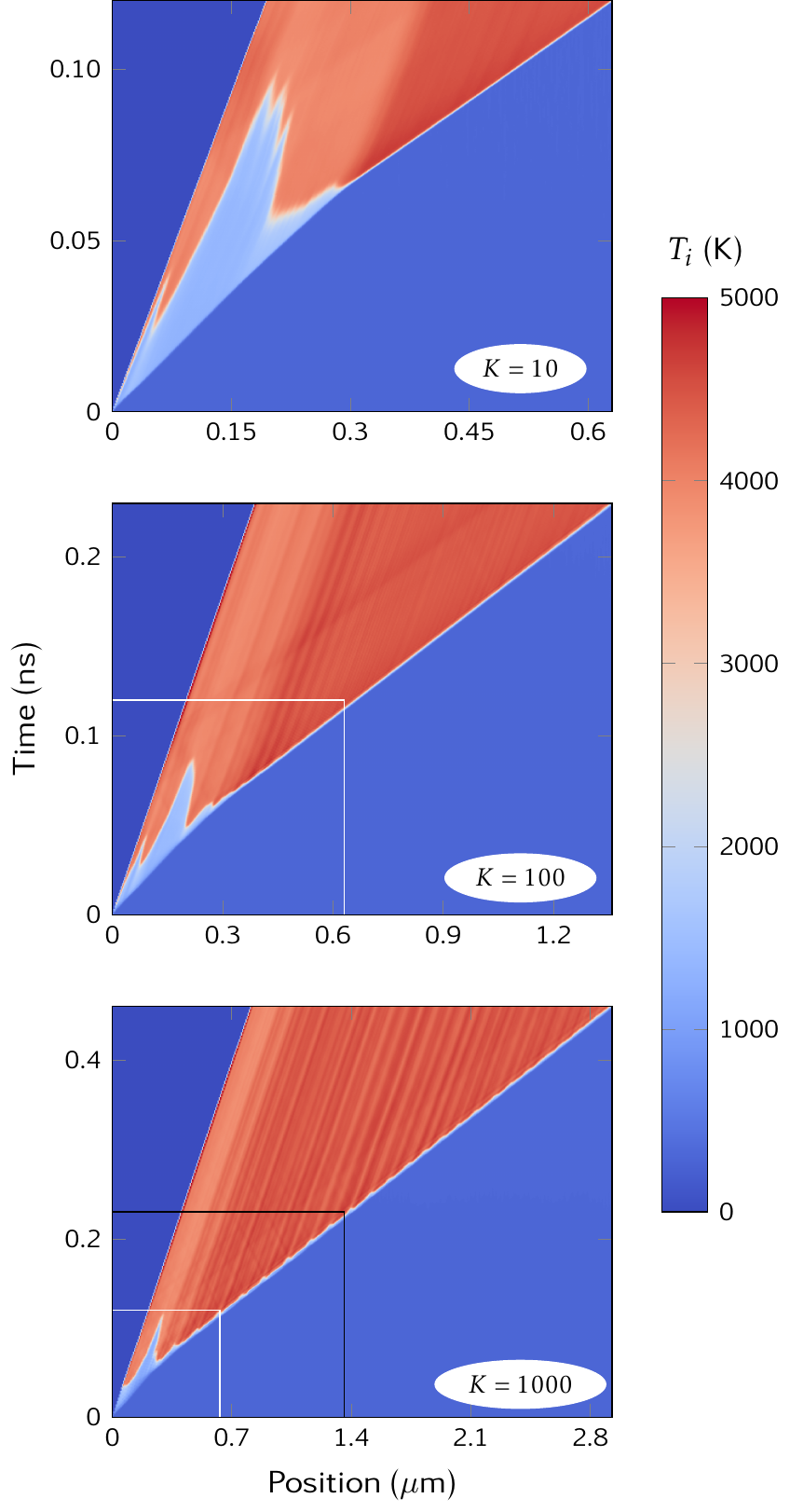}
  \caption{Space-time diagram of the temperature for several particle sizes: from $K=10$ to $K=1000$. The size of the space-time domain explored for each resolution is displayed in the diagram of the coarser simulations with a white ($K=10$) or black ($K=100$) frame.}
  \label{fig:color-deto-k}
\end{figure}

\begin{figure}[!ht]
  \centering
  \includegraphics{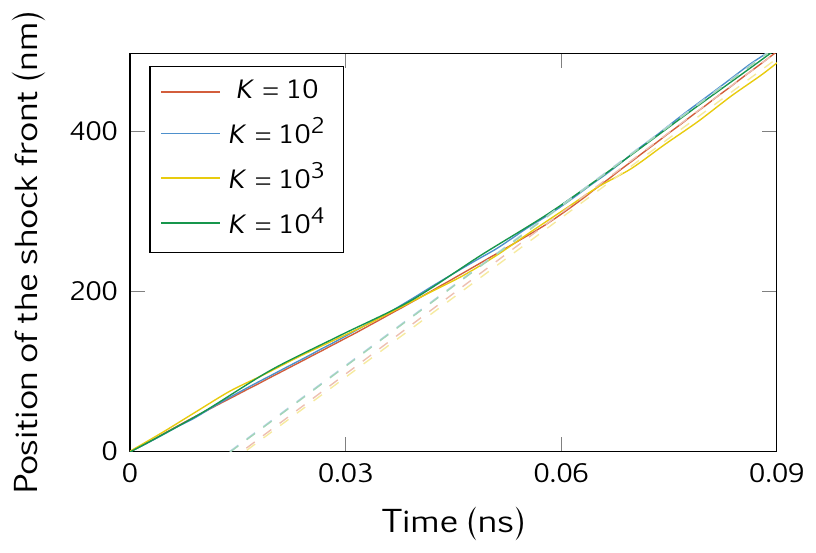}
  \caption{Position of the shock front with respect to time for different particle sizes. The linear interpolation during the detonation phase is plotted in dashed lines.}
  \label{fig:shockfront-k}
\end{figure}

We compare in Figure~\ref{fig:color-deto-k} the space-time diagram for particle sizes ranging from $K=10$ to $K=1000$.
We observe the same mechanism, with ignition points catching up with the shock front, at any resolution.
Moreover it seems that the shock-to-detonation transition occurs in the same physical time and length scales.
A more quantitative measurement of the invariance of the transition is to track the position of the shock front during its propagation (see Figure~\ref{fig:shockfront-k}).
Inside the two domains (the non reactive shock wave and the detonation wave), the shock front propagates with a constant velocity: $u_{\rm S}$ for the shock wave and $u_{\rm D}$ for the detonation wave.
The time to detonation is evaluated by extrapolating the linear evolution of the shock front in the detonation phase down to $t=0$.
The intercept of these linear interpolations with the time axis yields very close values for all particle sizes with a maximum $10\%$ difference between $K=10$ ($t_{\rm D}(z=0)=0.0157$~ns) and $K=10000$ ($t_{\rm D}(z=0)=0.0140$~ns).

\subsection{Steady detonation wave}
\label{sec:stationary}

We now slightly modify our setup to study steady detonation waves.
The system is still formed of $N=86400$ nitromethane particles on a $12\times12\times594$ grid at $\rho=1104$~kg.m$^{-3}$ and $T=300$~K.
A wall made of virtual SDPD particles~\cite{bian_2012,faure_2016} is placed at one end of the system.
At the other end, we insert a $50$~nm layer of nitromethane particles initialized at $\rho=1869$~kg.m$^{-3}$ and $T=2330$~K, which corresponds to the thermodynamic state obtained on the unreacted Hugoniot with a shock at $v_{\rm P} = 2500$~m.s$^{-1}$.
We observe a fast transition to a detonation wave that is followed by a rarefaction wave.

\begin{figure}[!ht]
  \centering
  \subfloat[]{\includegraphics{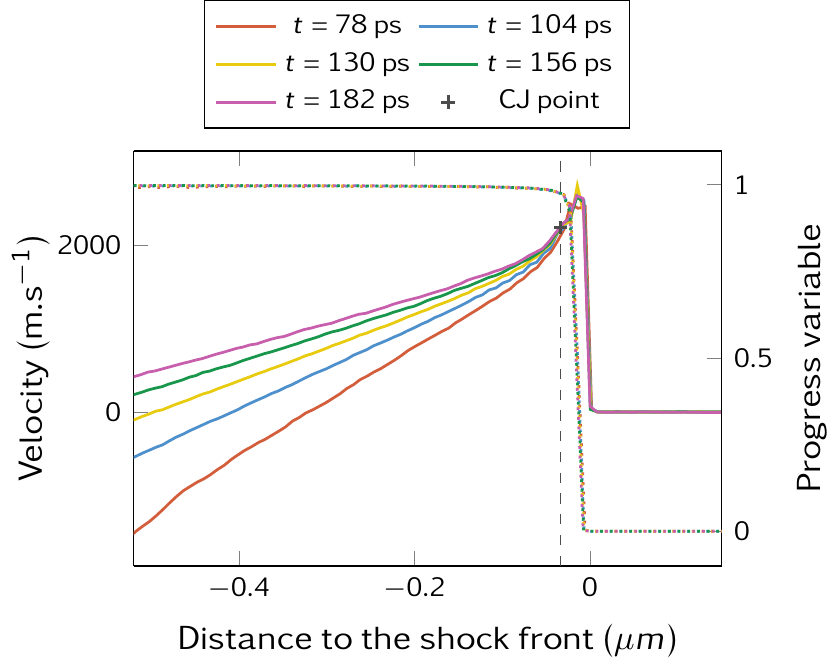}\label{fig:insta-velocity}}\\
  \subfloat[]{\includegraphics{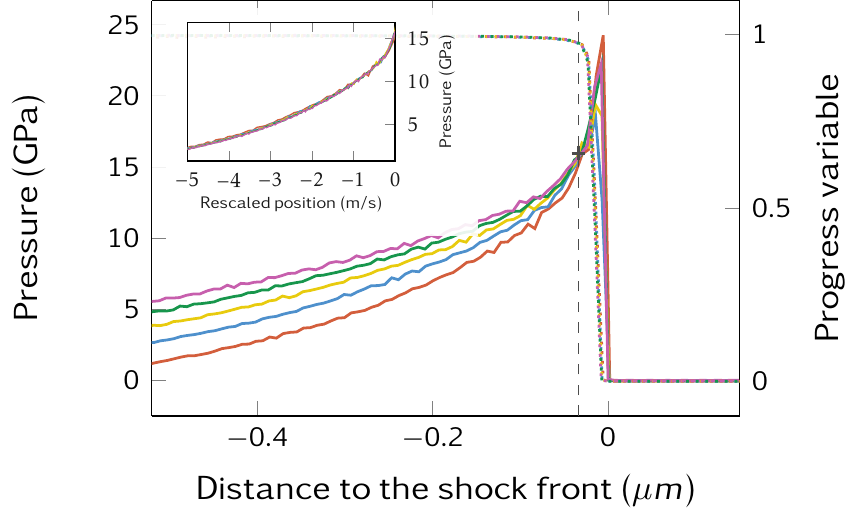}\label{fig:insta-pressure}}
  \caption{Instantaneous profiles of~\protect\subref{fig:insta-velocity} velocity and~\protect\subref{fig:insta-pressure} pressure in the reference frame of the shock front. The profile of the progress variable is also displayed in dotted line on both figures. The inset in~\protect\subref{fig:insta-pressure} is obtained by rescaling the positions as $\displaystyle \frac{z-z_{\rm CJ}}{t}$.}
  \label{fig:deto-insta}
\end{figure}

\begin{figure}
  \centering
  \includegraphics{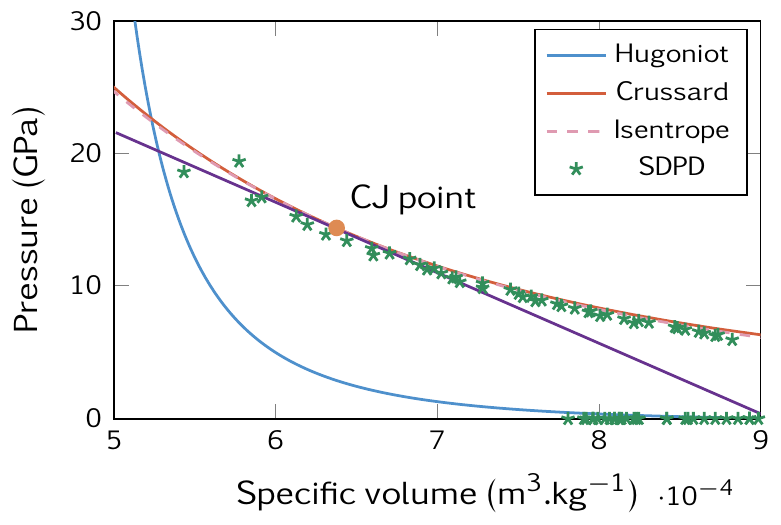}
  \caption{Hugoniot, Crussard and isentrope curve of nitromethane computed from the equations of state (solid lines) and the thermodynamic states observed in the SDPD simulations with $K=100$ (points). The Rayleigh line is also shown.}
  \label{fig:deto-pv}
\end{figure}

We check that we have reached the stationary regime with a reactive wave propagating at a constant velocity and a self-similar rarefaction wave.
Figure~\ref{fig:deto-insta} shows instantaneous profiles in the reference frame of the shock front for $K=100$.
\captionsetup[subfigure]{captionskip=-155pt}
The reactive zone, where the progress variable evolves from $0$ to $1$, is delimited by the pressure peak, just behind the shock front, and the CJ point determined in the pressure volume diagram (see Figure~\ref{fig:deto-pv}) as the tangent point between the Rayleigh line and the Crussard curve.
The rarefaction wave begins after the CJ point (located at position $z_{\rm CJ}$).
Upon rescaling the positions $z$ as $\displaystyle \frac{z-z_{\rm CJ}}{t}$ for each time $t$, the profiles coincide for $z<z_{\rm CJ}$, highlighting the self-similarity of the rarefaction wave.

In order to compare our results with theoretical predictions, we plot the instantaneous values of slice-averaged thermodynamic quantities in a pressure-volume diagram (see Figure~\ref{fig:deto-pv}) at time $t=130$~ps for $K=100$.
Up to the thermal fluctuations present in SDPD, the thermodynamic states observed in the rarefaction wave agree very well with the isentrope computed from the equation of state~(\ref{eq:eos-jwl}).
We summarize in Table~\ref{tab:deto-velocity} the detonation velocity for different particle size and compare them to the theoretical prediction obtained from the Rayleigh line in a simplified model that in particular does not account for viscosity effects.
\begin{table}[!ht]
  \centering
  \begin{tabular}{cc}
    \toprule
    Size & Detonation velocity (m.s$^{-1}$) \\
    \midrule
    Rayleigh & 6620\\
    10 & 6709\\
    100 & 6591\\
    1000 & 6549\\
    \bottomrule
  \end{tabular}
  \caption{Detonation velocity for different particle sizes compared to the theoretical hydrodynamic prediction.}
  \label{tab:deto-velocity}
\end{table}
The detonation velocities in SDPD are very close to the theoretical value.
Similarly to previous observations for shock waves~\cite{faure_2016}, the detonation velocity seems to be decreasing with the particle size.

\subsection{Influence of the Arrhenius prefactor}
\label{sec:prefactor}

\begin{figure}[!ht]
  \centering
  \includegraphics{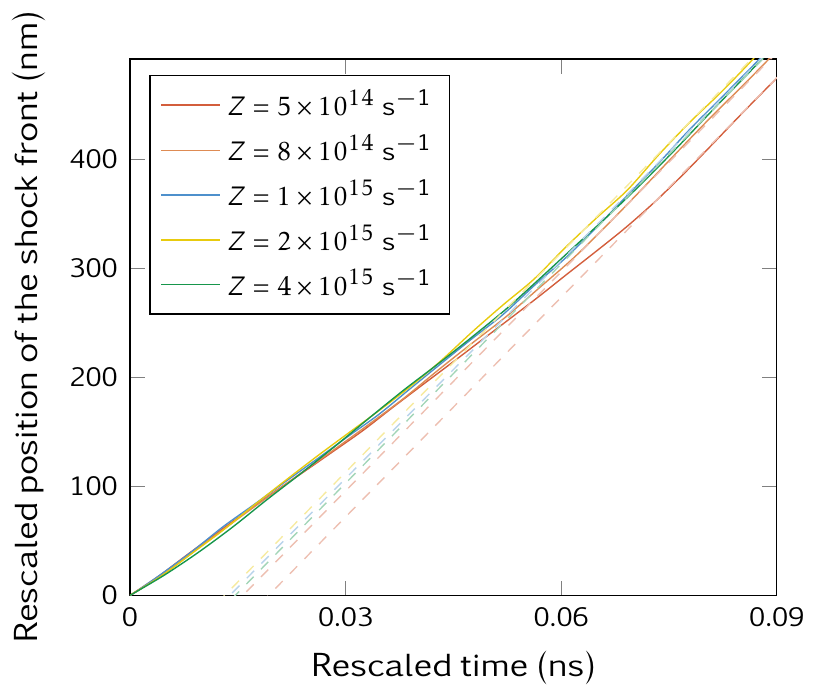}
  \caption{Position of the shock front with respect to time for different prefactors. }
  \label{fig:shockfront-z}
\end{figure}

All these results have been obtained with an arbitrarily chosen prefactor in the Arrhenius law~\eqref{eq:arrhenius}, namely $Z=10^{15}$~s$^{-1}$.
We study in the following the influence of the prefactor on the properties of the detonation wave.
We first turn our attention to the shock-to-detonation transition and perform the simulations reported in Section~\ref{sec:shock-to-deto} for $K=100$ and prefactors $Z=\num{5e14}$~s$^{-1}$, $Z=10^{15}$~s$^{-1}$ and $Z=\num{2e15}$~s$^{-1}$.
All the dimensions along the $z$-axis are scaled by a factor $\displaystyle \mathfrak{z} = \frac{Z}{10^{15}}$.
In all these settings the same mechanism is observed and we check whether a simple scaling law can predict the time to transition.
In Figure~\ref{fig:shockfront-z}, we plot the position of shock front with respect to time for several prefactors.
The positions and times are rescaled by the factor $\mathfrak{z}$.
Since we have taken $Z=10^{15}$~s$^{-1}$ as the reference, the scaling factor is $\mathfrak{z}=1$ for the prefactor $Z=10^{15}$~s$^{-1}$ and its profile remains unchanged in Figure~\ref{fig:shockfront-z} compared to Figure~\ref{fig:shockfront-k}.
It appears that, upon a simple rescaling, the trajectories of the shock front match reasonably well for the prefactors considered here, although the smallest prefactor ($Z=\num{5e14}$~s$^{-1}$) somewhat deviates from the others. 

\begin{figure}[!ht]
  \centering
  \includegraphics{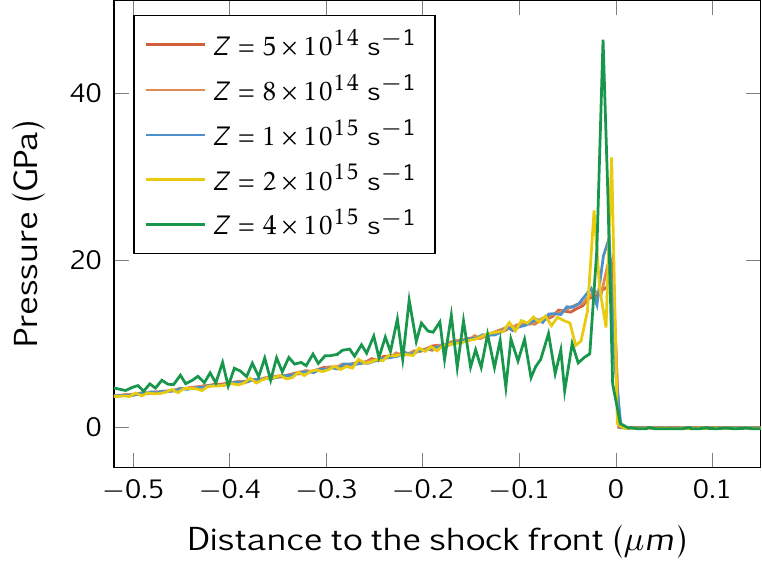}
  \caption{Pressure profiles for several prefactors at time $t=\frac{Z}{10^{15}}\times 130$~ps. The positions are rescaled with the same factor.}
  \label{fig:deto-profiles-z}
\end{figure}

We also study the influence of the Arrhenius prefactor on the stationary properties of the detonation wave and perform the simulations described in Section~\ref{sec:stationary} for $K=100$ and prefactors $Z=\num{5e14}$~s$^{-1}$, $Z=10^{15}$~s$^{-1}$ and $Z=\num{2e15}$~s$^{-1}$.
As for the STD transition, all the dimensions along the $z$-axis are scaled by the factor $\mathfrak{z}$.
We compare in Figure~\ref{fig:deto-profiles-z} the pressure profile at time $\mathfrak{z} \times 130$~ps for several prefactors.
The distances are also rescaled by the factor $\mathfrak{z}$.
The profiles agree very well with each other after the rescaling.
We notice however that as the prefactor increases higher pressures are observed at the shock front.
This can result in oscillations in the relaxation zone as clearly visible for $K=\num{4e15}$~s$^{-1}$.

In order to study the width of the reactive zone, we average the profiles of the progress variable over time.
We determine the reactive zone to be the region where the progress variable $\lambda$ is significantly different from $0$ or $1$, that is $0.02\leq\lambda\leq0.98$.
The widths for all tested prefactors along with the detonation velocity are gathered in Table~\ref{tab:deto-k}.
\begin{table}[!ht]
  \centering
  \begin{tabular}{ccc}
    \toprule
    Prefactor & Width of the & Detonation\\
    (s$^{-1}$) & reactive zone (nm) & velocity (m.s$^{-1}$)\\
    \midrule
    $\num{5e14}$ & $61.9$ & $6604$ \\
    $\num{8e14}$ & $39.9$ & $6599$ \\
    $\num{1e15}$ & $31.5$ & $6591$ \\
    $\num{2e15}$ & $15.6$ & $6622$ \\
    $\num{4e15}$ & $14.2$ & $7314$ \\
    \bottomrule
  \end{tabular}
  \caption{Width of the reaction zone and detonation velocity for several Arrhenius prefactors.}
  \label{tab:deto-k}
\end{table}
As expected, the detonation velocity remains very similar and close to the theoretical prediction which does not depend on the chemical kinetics.
As for the reactive region, its width roughly scales as $\mathfrak{z}^{-1}$, further suggesting that the prefactor simply involves a rescaling of the domain.

In Table~\ref{tab:deto-k}, it is manifest that the previous conclusions do not hold for $Z=\num{4e15}$~s$^{-1}$ for which even the detonation velocity is off by $15\%$.
Coupled with the observation in Figure~\ref{fig:deto-profiles-z}, this suggests that a finer resolution is required to deal with fast chemical reactions.
To confirm this, we run a simulation with the prefactor $Z=\num{4e15}$~s$^{-1}$ at a smaller particle size ($K=10$).
The detonation velocity determined with this setting, namely $u_{\rm D} = 6777$~m.s$^{-1}$, is much closer to the theoretical prediction.

While the change of prefactor in the regime explored by our simulations mainly amounts to rescaling the time and length scales, it appears that we should be careful to choose a sufficiently fine resolution.
As the kinetics are accelerated, with a larger prefactor, the reactive mechanism described in Section~\ref{sec:chemistry} in which the chemical reactions are averaged inside each mesoparticle, becomes unable to properly handle fast reactions for large particles.
Except for this limitation, SDPD has proved to be able to simulate detonation waves with a much coarser resolution than MD or DPDE and still recover not only the stationary properties but also the STD transition mechanism observed in~\cite{maillet_2007,maillet_2011}.

\section{Conclusion}
\label{sec:conclusion}

We have extended SDPD to handle reactive mechanisms.
This enables the simulation of detonation wave with SDPD.
The stationary properties, detonation velocity and thermodynamic states were succesfully recovered for nitromethane.
A mechanism with successive ignition points appearing in the shocked nitromethane was observed for the STD transition similarly to previous simulations with DPDE~\cite{maillet_2011}.
The resolution in SDPD has no major influence on these properties since the physical time and length scales associated with this process remain unchanged.
This allows us to effectively choose the resolution and deal with larger systems without affecting the physical properties of the detonation wave.
We have tested the influence of the prefactor on the STD transition and the stationary behavior of the reactive wave.
It actually seems that it only rescales the time and length scales in the simulation.
However it has stressed out that the resolution should be chosen fine enough with respect to the speed of the chemical reactions.
The multiscale consistency of reactive SDPD lets us envision to study more complex geometries by taking advantage of larger systems SDPD can simulate.

\section*{Acknowledgments}

We thank Nicolas Desbiens for fruitful discussions on the equations of state and the detonation mechanism. We are also grateful to Laurent Soulard and Gabriel Stoltz for their useful comments on the manuscript.

\end{document}